\documentclass[prb,showpacs,twocolumn,superscriptaddress]{revtex4}

\usepackage{graphicx}

\def\nbZ{{\mathchoice {\hbox{$\sf\textstyle Z\kern-0.4em Z$}}
{\hbox{$\sf\textstyle
Z\kern-0.4em Z$}} {\hbox{$\sf\scriptstyle Z\kern-0.3em Z$}}
{\hbox{$\sf\scriptscriptstyle Z\kern-0.2em Z$}}}}

\begin{document}

\title{Nonperturbative renormalization group approach to the Ising model: a
  derivative expansion at order $\partial^4$}

\author{L\'eonie Canet}
%\email{canet@lpthe.jussieu.fr}
\author{Bertrand Delamotte}
%\email{delamotte@lpthe.jussieu.fr}
\author{Dominique Mouhanna}
%\email{mouhanna@lpthe.jussieu.fr}

\address{Laboratoire de Physique Th\'{e}orique et Hautes \'Energies,
CNRS UMR 7589, \\
Universit\'{e} Pierre et Marie Curie Paris 6, Universit\'{e} Denis
Diderot Paris 7, 2
place Jussieu, 75252 Paris Cedex 05 France}
\author{Julien Vidal}
%\email{vidal@gps.jussieu.fr}
\address{Groupe de Physique des Solides, CNRS UMR 7588,\\
Universit\'{e} Pierre et Marie
Curie Paris 6, Universit\'{e} Denis Diderot Paris 7, 2 place Jussieu,
75251 Paris Cedex
05 France}

\begin{abstract}
  
On the example of the three-dimensional Ising model, we show that 
nonperturbative renormalization  group equations allow one to obtain very accurate
critical exponents. Implementing the order $\partial^4$  of the derivative expansion
leads to  $\nu=0.632$ and to an anomalous dimension $\eta=0.033$ which is
significantly improved compared with  lower orders calculations.

\end{abstract}

\pacs{05.10.Cc, 11.10.Gh, 11.10.Hi, 11.15.Tk, 64.60.-i}

\maketitle

%%%%%%%%%%%%%%%%%%%%%%%%%%%%%%%%%%%%%%%%%%%%%%%%%%%%%%%%%%%%%%%%%%%%%%%%%

Many problems  in high-energy as  well as in statistical  physics call
for nonperturbative methods. On the one hand, several physical systems
are described  by field  theories in their  strong coupling  regime so
that the  usual perturbative techniques become  troublesome. They fail
either because only the first  orders of perturbation are computed and
do not suffice, or because,  even when high orders are known, standard
resummation  techniques  do  not  provide
converged  results.  On the  other  hand,  some  phenomena such  as
confinement in QCD or phase transitions induced by topological defects
are genuinely nonperturbative.

Apart  from   some  methods  restricted  to   specific  dimensions  or
situations, very few nonperturbative techniques are available.  During
the   last  years,   the  Wilson   approach  \cite{wilson74}   to  the
renormalization  group (RG)  has been  turned into  an  efficient tool
\cite{wetterich93c,ellwanger94c,morris94a}.   This  nonperturbative RG
can be implemented  in very general situations and,  in particular, in
any  dimension, so that  it has  allowed one  to study  several issues
difficult   to   tackle   within  a  perturbative   framework among which  the
three-dimensional Gross-Neveu  model   \cite{hofling02},  frustrated
magnets  \cite{tissier}, the  randomly dilute  Ising model
\cite{tissier01b}, and the Abelian  Higgs model  \cite{bergerhoff96}.

This method  relies on a nonperturbative renormalization of the 
effective action
$\Gamma$, {\it i.e.} the Gibbs free energy. It consists in 
building an  effective
action $\Gamma_k$  at the  {\it running} scale  $k$  by integrating out only
fluctuations greater than $k$.  At the scale $k=\Lambda$, 
$\Lambda^{-1}$ denoting the
spacing of the  underlying  lattice, $\Gamma_k$ coincides with the 
Hamiltonian $H$
since {\it}  no  fluctuation has  yet been  taken into account while, at $k=0$, it
coincides with the standard effective action $\Gamma$ since   {\it all} fluctuations
have been integrated out. Thus, $\Gamma_k$   continuously 
interpolates   between   the
microscopic  Hamiltonian $H$ and the free energy $\Gamma$. The 
running effective  action
$\Gamma_k$ follows an exact equation which controls its evolution 
with the running scale
$k$ \cite{wetterich93c}:
%
%
%%%%%%%%%%%%%%%%%%%%%
\begin{equation}
\partial_t \Gamma_k[\phi] = {1 \over2} \int {d^dq\over (2\pi)^d}\
{\partial_t R}_k(q)
\left\{\Gamma_k^{(2)}[\phi(q)] + R_k(q)\right\}^{-1}
\mbox{,}
\label{eqflot}
\end{equation}
%%%%%%%%%%%%%%%%%%%%%
%
%
where  $t=\ln(k/\Lambda)$  and  $\Gamma_k^{(2)}[\phi]$ is  the
second functional
derivative  of  $  \Gamma_k$  with respect  to  the  field
$\phi(q)$. In Eq. (\ref{eqflot}), $R_k(q)$ is an  infrared cutoff 
function which
suppresses the propagation of the low-energy modes  without affecting 
the high-energy
ones.

Although   exact,   Eq.  (\ref{eqflot})   is   a  functional   partial
integro-differential  equation  which cannot  be  solved exactly.   To
handle it, one has to {\it truncate} $\Gamma_k$.  A natural and widely
used  truncation  is  the  derivative  expansion,  which  consists  in
expanding  $\Gamma_k$  in powers of $\partial  \phi$, keeping
only the lowest order terms.  Physically, this truncation rests on the
assumption that  the long-distance  physics of a  given model  is well
described   by  the   lowest   derivative  terms,   the  higher   ones
corresponding to less relevant  operators. Up to now, only truncations
up to  order $\partial^2$ have been  considered  since,
in many  cases, they  turn out  to be sufficient  to get  a satisfying
qualitative  and  even  sometimes  quantitative  description  of  both
universal and nonuniversal behaviors \cite{berges02}.

Nevertheless, several important issues concerning the reliability of
the method remain open. The  first one concerns the convergence of the
derivative expansion.  This point is  particularly delicate since,
within this  kind of  truncation, there is  no expansion  parameter in
terms of which the series  obtained can be analyzed and controlled. It
has  been  moreover  suggested   that  the  expansion  could  be  only
asymptotic \cite{bagnuls01}.  Actually, this question of  convergence has only
been    addressed    within     the    perturbative    context.     In
Refs. [\onlinecite{papenbrock95,morris99}],  it   has indeed been  shown  that  the
two-loop  perturbative result for  the  $O(N)$  model can  be recovered
from a  summation of  the derivative expansion.  However, in  its full
generality, this problem  still appears as  a major challenge. The
second  issue, the  truly interesting  one from  a practical  point of
view,  concerns the  accuracy of  the  results provided  by low orders
truncations. This  has been thoroughly studied
only  at order  $\partial^0$  within  the $O(N)$  model  and at  order
$\partial^2$ within  the Ising model  through the optimization  of the
cutoff function  $R_k$ \cite{litim02,canet03a} (see also related studies 
with the Polchinski equation \cite{ball95,comellas98} and  within the proper
time RG formalism \cite{zappala}).  
Let us emphasize that, even for these models,  the   anomalous  dimension  
$\eta$  remains  poorly
determined.  This  likely originates  in  the  crudeness  of the  order
$\partial^2$ truncation that fails to capture  the essential 
momentum  dependence  of
the  two-point correlation function. In this respect, an important 
remark is that in the
critical theory, and at $k=0$, this function  is  nonanalytic
---$\Gamma^{(2)}_{k=0}(q)\sim q^{2-\eta}$---, so it  appears 
nontrivial to  retrieve $\eta$ from a derivative expansion.  However, for $k\ne 0$, the   infrared
fluctuations    ($q\ll   k$)   are   suppressed   and 
$\Gamma^{(2)}_{k}(q)$  should  become
regular  with  the  standard  $q^2$ behavior.  This means that  the 
nonanalyticity
builds up  smoothly as $k$ vanishes. Berges {\it et al.} 
\cite{berges02}  have
proposed  that  this function behaves approximately as $q^2(q^2 +  c 
k^2)^{-\eta/2}$,
where $c$ is a constant.  Roughly  speaking,  for 
$\Gamma^{(2)}_{k}$,  the  derivative
expansion  consists in  expanding this  function around  $q=0$  and 
in computing $\eta$
from its behavior in  $k^2$, instead of $q^2$. It is not trivial 
that  the resulting
series  for $\eta$ converge since it amounts to  correct the normal 
$q^2$ behavior with
higher  powers of $q^2$.
 The aim of this paper is to investigate this question by 
including order $\partial^4$
terms in the derivative expansion of $\Gamma_k$ for the 
three-dimensional Ising model.

The effective average action $\Gamma_k$ of the Ising model truncated at
order $\partial^4$ is written as:
%
%
%%%%%%%%%%%%%%%%%%%
\begin{eqnarray}
\Gamma_k[\phi]&=&\int d^d x\:\Bigg\{U_k\left(\rho \right) +
\nonumber\\ && {1\over 2}
\Big[ Z_k(\rho)\left(\nabla\phi\right)^2 +
W^a_k(\rho)\left(\Delta\phi\right)^2 +\\ &&
W^b_k(\rho)\left(\nabla\phi\right)^2 \left(\phi\Delta\phi \right) +
W^c_k(\rho)\left(\left(\nabla\phi\right)^2\right)^2
\Big] \Bigg\},\nonumber
\label{derivexp}
\end{eqnarray}
%%%%%%%%%%%%%%%%%%%
%
%
where  $\rho=\phi^2/2$ is  the $\nbZ_2$  invariant. Compared  with its
expansion at order  $\partial^2$,  $\Gamma_k$  involves  three new 
terms denoted
$W_k^s(\rho)$ $s=a,b,c$, linearly  independent with respect to the
integration  by parts.  The  evolution equation  for the  potential $U_k$ is
derived   by  evaluating  Eq. (\ref{eqflot})   for  a  uniform  field
configuration. By contrast,
the  definition and,  thus, the  evolution  of the  functions $Z_k$  and
$W^s_k$'s are  linked  to   a  specific  momentum  dependence  of  the
functional  derivatives  of  $\Gamma_k$,  in the  limit  of  vanishing
external momenta:
%
%
%%%%%%%%%%%%%%%%%%
\begin{eqnarray}
Z_k(\rho) &=& \lim_{p_i\rightarrow 0} \partial_{p_1^2}
\frac{\delta^2 \Gamma_k}{\delta\phi(p_1)\delta\phi(p_2)},
\label{definitionsz}\\
W^a_k(\rho) &=& \lim_{p_i\rightarrow 0} \partial_{p_1^4}
\frac{\delta^2 \Gamma_k}{\delta\phi(p_1)\delta\phi(p_2)},
\label{definitionswa}\\
W^b_k(\rho) &=&-\frac{1}{2\sqrt{2 \rho}} \lim_{p_i\rightarrow 0} 
\partial_{p_1^2 p_2^2}
\frac{\delta^3 \Gamma_k}{\delta\phi(p_1)\delta\phi(p_2)\delta\phi(p_3)},
\label{definitionswb}\\
W^c_k(\rho) &=&-\frac{1}{4} \lim_{p_i\rightarrow 0}
\partial_{p_1^2\vec{p}_2.\vec{p}_3}
\frac{\delta^ 4
\Gamma_k}{\delta\phi(p_1)\delta\phi(p_2)\delta\phi(p_3)\delta\phi(p_4)}
\mbox{.}
\nonumber\\
\label{definitionswc}
\end{eqnarray}
%%%%%%%%%%%%%%%%%%
%
%
As usual, to find a fixed point, we  use  the associated 
dimensionless renormalized
quantities  $\bar{\rho},u_k$, $z_k$ and $w_k^s$. The flow equations of these
functions, derived from  Eq. (\ref{eqflot}), are far too long to be 
displayed.

As in Ref. [\onlinecite{canet03a}], we have  implemented a further
approximation
which consists in expanding each running function $u_k$,
$z_k$ and the $w^s_k$'s in powers of $\bar{\rho}$. The  motivation 
which underlies
this is twofold.
First, in systems having a symmetry group smaller than $O(N)$, the
number of functions
analogous to $u_k$,
$z_k$ and $w^s_k$ grows as well as the number of arguments, analogous to
$\bar{\rho}$, on which they depend. In this case,  dealing with the full
field dependence at each order of the derivative expansion
can be very  demanding and the  field expansion becomes almost
unavoidable. Second, this expansion provides valuable indications about the orders
in field necessary to correctly describe the critical behavior. 
Here, we  expand the $u_k$, $z_k$ and $w^s_k$ functions around the configuration
$\bar{\rho}=\bar{\rho}_0$  that minimizes
$u_k$ since it  leads to a better convergence than the expansion
around $\bar{\rho}=0$ \cite{aoki98}:
%
%
%%%%%%%%%%%%%%%%%%%
\begin{eqnarray}
\zeta_k &=&\displaystyle{ \sum_{j=0}^{\mbox{$p_{\mbox{\tiny $\zeta$}}$}}
\zeta_{j,k} (\bar{\rho}-\bar{\rho}_0)^j}
\mbox{,}
\label{z0}
\end{eqnarray}
%%%%%%%%%%%%%%%%%%%
%
%
where $\zeta$ stands for $u,z,w^a,w^b,w^c$. The RG equation
(\ref{eqflot}) then leads to
a set of ordinary  coupled differential equations for the coupling constants
$\left\{\zeta_{j,k}\right\}$. The
nonperturbative features of
their flows with the  running scale $k$ are entirely encoded in a finite set of
integrals, called threshold functions. There are six --  three 
of them being specifically  linked to the inclusion of the $\partial^4$ order terms
-- which are written:
%
%
%%%%%%%%%%%%%%%%%%%
\begin{equation}
F_n^d =\int  dyy^{\frac{d}{2}-1}
 {\tilde \partial_t}\bigg(f(y)\frac{1}{(p(y)+ m^2)^n}\bigg)
\mbox{,}
\label{thresholds}
\end{equation}
%%%%%%%%%%%%%%%%%%%
%
%
where  ${\tilde \partial_t}$ means that the derivative only acts on the cutoff
function $R_k(q) = Z_{0,k} \:q^2  r(y)$ with $y=q^2/k^2$; $p(y)= y(1 +
w^a_{0,k} \: y + r(y))$, $m^2= 2 u_{2,k} \:\bar\rho_0$, and $f(y)$ can be
either  
$y(\partial_y p)^i$ with $i=0,...,4$ or $y \partial^2_y p$.  The occurence
of $\partial^2_y p$   imposes the cutoff function $R_k$ to be at least of
class $C^3$, which dismisses for instance  the theta cutoff introduced in 
\cite{litim01}. Here, we choose the exponential cutoff defined by
\cite{wetterich93c}:
%
%
%%%%%%%%%%%%%%%%%%%
\begin{equation}
R_k(q) = \alpha \frac{Z_{0,k}\;q^2}{e^{\:q^2/k^2}-1}
\mbox{,}
\label{cutoff}
\end{equation}
%%%%%%%%%%%%%%%%%%%
%
%
which fulfills this condition and constitutes an efficient regulator.
We remind that any truncation of $\Gamma_k$ introduces a spurious dependence of
the results on $R_k$. Here, we study this influence by varying the cut-off
through the amplitude parameter $\alpha$. 
For each truncation, the optimal $\alpha$ is determined through a principle of
minimum sensitivity  (PMS) which indeed corresponds to an optimization of the
accuracy of the critical exponents \cite{canet03a}.

%
%
%%%%%%%%%%%%%%%%%%
\begin{figure}[h]
\includegraphics[width=75mm,height=90mm]{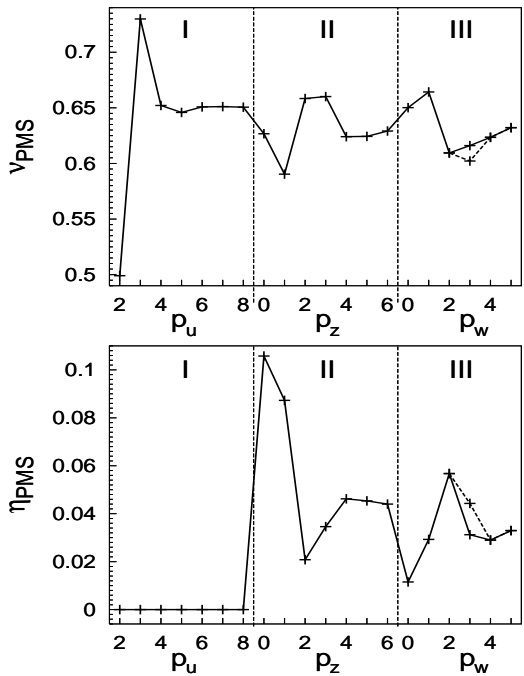}
\caption{$\nu_{\hbox{\tiny PMS}}$ and $\eta_{\hbox{\tiny PMS}}$ as functions of 
the truncation. The
three zones I, II and III correspond to the expansions of $u_k$, 
$z_k$ and the $w_k^s$'s
respectively. In zone III, the two values at $p_{w^{s}}=3$ reflect the
different choices of PMS solutions (see below).}
\label{cv1}
\end{figure}
%%%%%%%%%%%%%%%%%%
%
%

At each order  of the derivative expansion, up  to order $\partial^4$,
and for  higher and higher  order field truncations, we compute the
fixed point and the associated critical exponents $\nu$ and $\eta$, as
functions of  $\alpha$.  Then, for  each truncation, we  determine the
optimized exponents from the PMS, which are referred to, in the following,
as PMS exponents.  We first  expand in fields the potential $u_k$, and
then $z_k$, which respectively  constitute the orders $\partial^0$ and
$\partial^2$  of  the  derivative  expansion.  The  corresponding  PMS
exponents are displayed as functions of $p_{u,z}$ --- which denote the
orders of the truncation in $\bar{\rho}$ of $u_k$ and $z_k$ --- in the
first  two  zones of  Fig.   \ref{cv1}. At  this  stage,  it is  worth
emphasizing that strong oscillations occur  at the first orders in the
field  expansion for  both orders  $\partial^0$ and  $\partial^2$.  It
follows that the  PMS exponents  become almost steady  only from $p_{u,z}= 4$.
As  discussed  in  Ref. [\onlinecite{canet03a}],  the  truncation  $p_u=8$  and
$p_z=6$ allows one to obtain a  very accurate approximation of the order
$\partial^2$ results. Indeed,  the corresponding exponents 
$\nu_{\hbox{\tiny PMS}}=0.6291$ and  $\eta_{\hbox{\tiny PMS}}=0.0440$ differ by  less than $1\%$
compared with their ``asymptotic'' values obtained for large $p_{u,z}$
(see  Table  \ref{tab}).   Note  also  that,  already  at  this  order
$\partial^2$, $\nu_{\hbox{\tiny PMS}}$ agrees well with the best known values
whereas, as mentioned above, this is not the case for $\eta_{\hbox{\tiny PMS}}$.

\begin{figure}[ht]
\includegraphics[width=50mm,height=90mm,angle=-90]{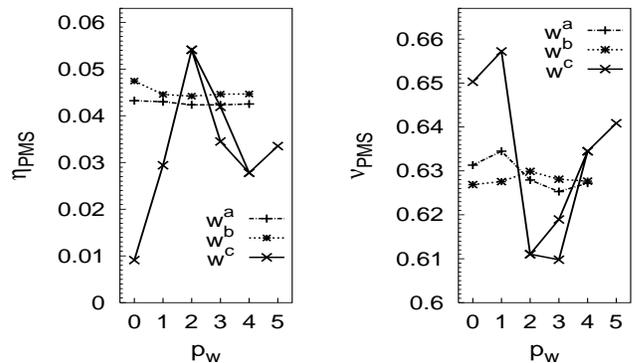}
\caption{$\nu_{\hbox{\tiny PMS}}$ and $\eta_{\hbox{\tiny PMS}}$ as  
functions of the order of the field  truncation for each function $w^s_k$ 
separately. For $w^c$, the two values at $p_{w^{c}}=3$
reflect the different choices of PMS solutions (see below). }
\label{cv2}
\end{figure}

Let us come to the role  of the order $\partial^4$ terms. We choose to
simultaneously   expand  in  fields   the  three   functions  $w^s_k$,
$s=a,b,c$,   up    to   $p_{w^s}=5$,   while    fixing   $p_u=8$   and
$p_z=6$. Actually, the  highest truncation corresponds to $p_{w^c}=5$
and  $p_{w^{a,b}}=4$   for  the  following   reason.   Figure  \ref{cv2}
displays, for each $w^s_k$ considered independently, the evolutions of
the PMS exponents with the order of the field truncation.
%
%
%%%%%%%%%%%%%%%%%%
%%%%%%%%%%%%%%%%%%
%
%
It shows  that the exponents  associated with $w^a_k$ or  $w^b_k$ have
almost  converged,  up  to  a  few  tenths  of  percent,  as  soon  as
$p_{w^{a,b}}=3$.  On  the   contrary,  $\eta_{\hbox{\tiny  PMS}}$  and
$\nu_{\hbox{\tiny PMS}}$ related to $w^c_k$ still oscillate at this order. We
have  checked that,  within the  simultaneous expansion  of  the three
$w^s_k$,  $w^c_k$  indeed  dictates  the variations  of  the  critical
exponents, $w_{j,k}^{a}$ and $w_{j,k}^{b}$ exerting a minor influence for $j\ge
3$. 
 
This, together with the fact  that we encounter here the limits  of our computational 
capacities, justifies our  choice
$(p_{w^{a}},p_{w^{b}},p_{w^{c}})=(4,4,5)$ for the last truncation.

\begin{figure}[h]
\includegraphics[width=75mm,height=90mm]{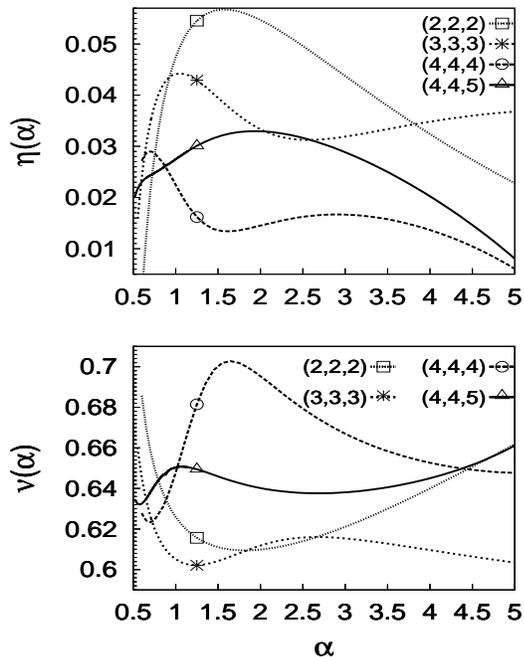}
\caption{Variations of the critical exponents $\nu$ and $\eta$ with the
cutoff parameter
$\alpha$, Eq. (\ref{cutoff}), for the highest orders truncations labelled by
$(p_{\mbox{\tiny $w^a$}},p_{\mbox{\tiny $w^b$}},p_{\mbox{\tiny $w^c$}})$.}
\label{PMS}
\end{figure}

  We can now concentrate on the behavior of the exponents at the order $\partial^4$.
At the low orders field truncations,
corresponding to  $p_{w^s}=0,1,$ and $2$, each exponent exhibits a
single PMS solution,
$\nu_{\hbox{\tiny PMS}}$ and $\eta_{\hbox{\tiny PMS}}$, which are thus unambiguously
defined. As displayed in Fig. \ref{PMS}, several PMS solutions appear for
the next two  truncations, corresponding to $p_{w^s}=3,4$. This
renders the optimization procedure in these cases (see discussion below) unclear. 
%
%
%%%%%%%%%%%%%%%%%%
%%%%%%%%%%%%%%%%%%
%
%
Concerning  the  largest   truncation,  $\eta_{\hbox{\tiny  PMS}}$  is
unambiguously  determined from  the unique  PMS solution.   For $\nu$,
several PMS  solutions exist.   However, provided the  field expansion
has almost converged at this order,  a unique PMS solution can also be
selected for  $\nu$. The argument underlying this choice originates from the
fact that when no truncation in derivatives is performed, 
the results are independent of the cutoff.
Therefore, the best cutoff is the one achieving the weakest sensitivity of the
results with respect to the order of the derivative expansion, {\it i.e.} leading
to the fastest convergence (see Ref. [\onlinecite{canet03a}] for a detailed
discussion).  In practice, this consists in  minimizing the difference  between
the values of 
$\nu$  determined  at  order  $\partial^n$  and at order $\partial^{n+1}$.  In our
case, this selects the PMS solution located at $\alpha\simeq 0.6$ (see Fig.
\ref{PMS}).

Let us now discuss the convergence  of the field expansion. To this end,
 we first  examine the two truncations $p_{w^{s}}=3,4$  for which multiple
 PMS  solutions exist  for both  exponents.  There  is no  argument to
 clearly settle between the PMS solutions.  We present two
 sensible way  to favor reasonable PMS solutions. 
 First, one  can choose to minimize, for  these orders, the
 oscillations  induced   by  the  field  expansion.   This,  in  turn,
 corresponds  to improving the  rapidity of  convergence of  the field
 expansion. This choice corresponds to the full line in the third zone
 of Fig.  \ref{cv1}. Alternatively, one  can decide to follow  a given
 PMS  solution  (characterized  by  its  concavity  and  its  location
 $\alpha$), order by  order in the field expansion,  starting from the
 low orders  $p_{w^{s}}=0,1,2$, where the  PMS solutions are  unique. This
 corresponds to the dashed line in Fig.  \ref{cv1}.  Note that both criteria   
lead   to    the   same   PMS   solution   for
 $p_{w^{s}}=4$. Finally, the important features of the exponents evolution
 remain essentially  unchanged whatever choice is  adopted: low orders
 generate strong oscillations that tend  to vanish after a few orders.
 Indeed,  the results  for $p_{w^{s}}=4$:  $\nu_{\hbox{\tiny PMS}}=0.6234$
 and  $\eta_{\hbox{\tiny PMS}}=0.0289$,  are very  close to  those for
 $p_{w^{s}}=5$:  $\nu_{\hbox{\tiny  PMS}}=0.6321$  and  $\eta_{\hbox{\tiny
 PMS}}=0.0330$.   Although the  exponents are  not  rigorously steady,
 this suggests  that the asymptotic  regime is just entered.   This is
 consistent   with  the   fact  that,   at  orders   $\partial^0$  and
 $\partial^2$,  the  oscillations  die  down  for  the  same  order  of
 truncation:   $p_{u,z}\simeq  4  $.   This  legitimates   our  former
 assumption of  field convergence. We therefore  approximate the order
 $\partial^4$ results by the $p_{w^{s}}=5$ estimates, see Table I.
%
%
%%%%%%%%%%%%%%%%%%%%%%%%%%%%%%%%%%%%%%%%%%%%%%55
\begin{table}[b]
\begin{tabular}{|l|c|c|}
\hline
   method             & $\nu$  & $\eta$    \\
\hline
LPA$^{(a)}$           & 0.6506   &  0       \\
\hline
$\partial^{2 \;(a)}$  & 0.6281  &  0.0443  \\
\hline
$\partial^{4 \;(b)}$  & 0.632   &  0.033  \\
\hline
7-loop$^{(c)}$        & 0.6304(13)& 0.0335(25)\\
\hline
MC$^{(d)}$            & 0.6297(5) & 0.0362(8)\\
\hline
\end{tabular}
\caption{Critical exponents of the three-dimensional Ising model:
$a$) effective average action method (field expansion)
\cite{canet03a}; $b$) present work;
$c$) 7-loop calculations \cite{guida98}; $d$) Monte-Carlo simulations
\cite{hasenbusch01}.}
\label{tab}
\end{table}
%%%%%%%%%%%%%%%%%%%%%%%%%%%%%%%%%%%%%%%%%%%%%%
%
%
To summarize, we have computed the critical exponents of the
three-dimensional Ising model up to the $\partial^4$ order in the derivative
expansion. The successive contributions significantly decrease with the order,
which supports good convergence properties of this expansion, and in particular a
correct behavior of its implementation around $q=0$. We emphasize that 
the  exponent $\nu$ is  almost unaltered at order $\partial^4$ compared with its 
value at  the order $\partial^2$,  whereas $\eta$ undergoes  a  substantial
correction  which  drives  it  within a  few percents of the best known values.  
This confirms the statement that the inclusion of the $\partial^4$ order terms
allow one to improve the anomalous dimension. 
Note that  although  fully converged results would require to handle the full
field dependence of $u_k$, $z_k$ and the  $w_k^s$'s \cite{canet03c},  
this study shows that the truncation in fields constitutes a reliable way to
compute critical exponents. Finally, the present work brings out convincing
evidence of the ability of the effective average action method to provide 
very accurate estimates of physical quantities.

%\bibliography{Dominique}

\begin{thebibliography}{10}

\bibitem{wilson74}
K.~G. Wilson and J. Kogut, Phys. Rep. C, Phys. Lett. {\bf 12},  75  (1974).

\bibitem{wetterich93c}
C. Wetterich, Phys. Lett. B {\bf 301},  90  (1993).

\bibitem{ellwanger94c}
U. Ellwanger, Z. Phys. C {\bf 62},  503  (1994).

\bibitem{morris94a}
T.~R. Morris, Int. J. Mod. Phys. A {\bf 9},  2411  (1994).

\bibitem{hofling02}
F. H{\"o}fling, C. Nowak, and C. Wetterich, Phys. Rev. B {\bf 66},  205111
  (2002).

\bibitem{tissier}
M. Tissier, B. Delamotte, and D. Mouhanna, Phys. Rev. Lett. {\bf 84},  5208
  (2000), Phys. Rev. B {\bf 67},  134422 (2003).

\bibitem{tissier01b}
M.Tissier, D. Mouhanna, J. Vidal, and B. Delamotte, Phys. Rev. B {\bf 65},
  140402  (2002).

\bibitem{bergerhoff96}
B. Bergerhoff {\it et~al.}, Phys. Rev. B {\bf 53},  5734  (1996).

\bibitem{berges02}
J. Berges, N. Tetradis, and C. Wetterich, Phys. Rep. {\bf 363},  223  (2002).

\bibitem{bagnuls01}
C. Bagnuls and C. Bervillier, Phys. Rep. {\bf 348},  91  (2001).

\bibitem{papenbrock95}
T. Papenbrock and C. Wetterich, Z. Phys. C {\bf 65},  519  (1995).

\bibitem{morris99}
T.~R. Morris and J.~F. Tighe, JHEP {\bf 08},  007  (1999).

\bibitem{litim02}
D. Litim, Nucl. Phys. B {\bf 631},  128  (2002).

\bibitem{canet03a}
L. Canet, B. Delamotte, D. Mouhanna, and J. Vidal, Phys. Rev. D {\bf 67},
  065004  (2003).

\bibitem{ball95}
R.~D. Ball, P.~E. Haagensen, J.~I. Latorre, and E. Moreno, Phys. Lett. B {\bf
  347},  80  (1995).

\bibitem{comellas98}
J. Comellas, Nucl. Phys. B {\bf 509},  662  (1998).

\bibitem{zappala}
M. Mazza and D. Zappala, Phys. Rev. D {\bf 64},  105013  (2001).

\bibitem{aoki98}
K.~I. Aoki {\it et~al.}, Prog. Theor. Phys. {\bf 99},  451  (1998).

\bibitem{litim01}
D. Litim, Phys. Rev. D {\bf 64},  105007  (2001).

\bibitem{guida98}
R. Guida and J. Zinn-Justin, J. Phys. A {\bf 31},  8103  (1998).

\bibitem{hasenbusch01}
M. Hasenbusch, Int. J. Mod. Phys. C {\bf 12},  911  (2001).

\bibitem{canet03c}
L. Canet, B. Delamotte, D. Mouhanna, and J. Vidal, (unpublished).

\end{thebibliography}
%\bibliographystyle{prsty}

\end{document}